\documentclass[review]{elsarticle}

\usepackage{hyperref,graphicx,wrapfig,float,textcomp}
\graphicspath{{images/}}

\journal{Journal of \LaTeX\ Templates}

\begin{document}

\begin{frontmatter}

\title{Driving Digital Rock towards Machine Learning: predicting permeability with Gradient Boosting and Deep Neural Networks}

\author[addressa]{Oleg Sudakov\footnote{Dataset generation, model implementation, model evaluation}}
\author[addressb]{Evgeny Burnaev\footnote{Problem formulation, machine learning consulting, scientific supervision}}
\author[addressc]{Dmitry Koroteev\footnote{Problem formulation, petrophysics consulting, scientific supervision}}

\address[addressa]{Skolkovo Institute of Science and Technology, Skolkovo Innovation Center, Building 3, Moscow, 143026, Russia, oleg.sudakov@skolkovotech.ru}
\address[addressb]{Skolkovo Institute of Science and Technology, Skolkovo Innovation Center, Building 3, Moscow, 143026, Russia, e.burnaev@skoltech.ru}
\address[addressc]{Skolkovo Institute of Science and Technology, Skolkovo Innovation Center, Building 3, Moscow, 143026, Russia, d.koroteev@skoltech.ru}

\begin{abstract}
We present a research study aimed at testing of applicability of machine learning techniques for prediction of permeability of digitized rock samples. We prepare a training set containing 3D images of sandstone samples imaged with X-ray microtomography and corresponding permeability values simulated with Pore Network approach. We also use Minkowski functionals and Deep Learning-based descriptors of 3D images and 2D slices as input features for predictive model training and prediction. We compare predictive power of various feature sets and methods. The later include 
Gradient Boosting and various architectures of Deep Neural Networks (DNN). The results demonstrate applicability of machine learning for image-based permeability prediction and open a new area of Digital Rock research.  

\end{abstract}

\begin{keyword}
\texttt {Digital Rock \sep Machine Learning \sep Artificial Neural Networks \sep permeability prediction \sep gradient boosting}
\end{keyword}

\end{frontmatter}

\section{Introduction and Justification}

“Digital Rock Physics” is an innovative approach for computing the properties of rocks. The paradigm of Digital Rock Physics is “image-and-compute”: the rock sample is imaged to obtain a 3D representation of the mineral phase and pore space, and this 3D representation is then used to simulate physical processes in the sample. \cite{chauhan2016processing, koroteev2017method, andra2013digital_1,blunt2013pore}. 

Recent methods of 3D imaging of pore topology include micro-scale x-ray computed tomography (\textmu xCT), which images rock samples with resolution down to tens of nanometers (voxel size) or hundreds of nanometers (physical resolution). \textmu xCT enables the internal structure of fine-structured samples to be imaged accurately and non-destructively. After removal of \textmu xCT scanning artifacts and segmentation \cite{chauhan2016processing, koroteev2017method}, the scan is processed to retain the sample’s petrophysical properties.

Applications of Digital Rock technologies include:

\begin{itemize}

\item the calculation of transport properties such as absolute permeability and relative permeability \cite{andra2013digital_2, koroteev2014direct, berg2017industrial, blunt2013pore};

\item the calculation of electric, elastic, geomechanical properties and NMR response \cite{andra2013digital_2, blunt2013pore, evseev2015coupling};

\item screening enhanced oil recovery methods \cite{koroteev2013application};

\item assessing the potential efficiency of chemical treatment for well stimulation \cite{klemin2015digital}.

\end{itemize}

Recent advances in high-resolution imaging, high performance computing and Machine Learning will lead to new and more effective computation.

The present work applies advances in Deep Learning image processing to Digital Rock Physics. Our goal is to build fast approximation models, or so-called surrogate models, to predict permeability based on the results of physical modeling (an example of such modeling can be found in \cite{belyaev2016gtapprox}). Such models are an acknowledged method for solving various industrial engineering problems \cite{grihon2013surrogate}. 

Using the VGG-16 DNN \cite{simonyan2014very} network, we recover a set of descriptors for the 2D layers, which compose the 3D image, and utilize their low-dimensional representation to compute sample permeability. The results outperform the frequently used technique of using Minkowski functionals as input features for a machine learning algorithm to predict logarithmic permeability.

We also assess the applicability of conventional Deep Learning models — convolutional neural networks (CNNs), which are frequently used in the analysis of multidimensional data — to \textmu xCT voxel rock scans. We apply CNNs in an end-to-end fashion to simultaneously extract features and carry out regression for permeability prediction. The advantage of this approach is that it does not require manual feature engineering, but provides equally accurate results.

\section{Data Acquisition}

We used a sample from the Berea Sandstone Petroleum Cores (Ohio, USA) for model evaluation. The 3D image already had its artifacts removed and its segmentation computed by Imperial College London. The segmented sample makes no distinction between different rock phases, denoting every rock voxel as 0, and every pore voxel as 1. The initial sample consisted of $400 \times 400 \times 400$ elements with voxel size of 5.345 \textmu m.

\begin{wrapfigure}{r}{0.5\textwidth}
    \centering  \includegraphics[width=0.4\textwidth]{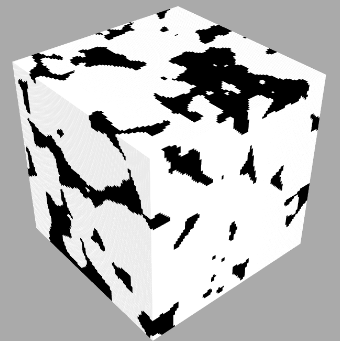}
    \caption{Berea sandstone sample}
\end{wrapfigure}

To generate a dataset for machine learning algorithms, the sample was cut into intersecting $100 \times 100 \times 100$ voxel cubes with shift of 15 voxels and same step size of 5.345 \textmu m, giving a dataset of 9261 samples in total. Each of these smaller samples can be examined as an independent rock image, retaining some geometrical properties of the parent Berea sample.

To compute initial permeability values for voxel cubes in the dataset, we used a pore-scale network modelling code (courtesy of Taha Sochi) \cite{sochi2010pore}, paired with OpenPNM framework \cite{putz2013introducing}. The network model is a simplified representation of rock geometry, consisting of spherical pores connected by cylindrical throats, usually stored in Statoil format. The network representation was then used to compute the rock permeability of each cut rock sample, making use of Darcy’s law and taking the flow type to be Stokes flow. The result was a dataset of 9261 $100 \times 100 \times 100$ voxel cubes and corresponding permeability values, measured in millidarcies. The by-product of the calculations is a set of 9261 network models.

\section{Regression on Generated Features}

\subsection{Feature generation}

We considered three different approaches to feature generation for regression. First, we tried to explicitly use characteristics of network models. Second, we computed a well-known set of geometrical descriptors, Minkowski Functionals, for use as an input for the predictor. Finally, we used a set of 2D image descriptors with reduced dimensionality, acquired using VGG-16 DNN and Principal Component Analysis (PCA), as a feature set.

\paragraph{Network characteristics}

We considered a number of network model characteristics, which could influence the permeability value for a given sample. These characteristics are: median pore radius, mean pore radius, median throat radius, mean throat radius, median throat length, mean throat length, median pore connectivity number, mean pore connectivity number, and total pore count.
As in Stokes flow, we considered advective inertial forces to be small, compared to viscous forces. The permeability of the sample is then proportional to the area of the phase transition surface, which, in turn, is proportional to certain included characteristics.
However, this approach proved inferior to the other two approaches.

\paragraph{Minkowski functionals}

Minkowski functionals (also known as intrinsic volumes or quermass integrals) are additive morphological measures, initially defined for convex objects in the field of integral geometry. It has been shown that every measure on the finite union of compact convex sets in $R^3$ can be expressed as a linear combination of the four Minkowski functionals. For $R^3$ these Minkowski functionals are volume, area, mean breadth and the Euler-Poincaré characteristic. In recent years, these functionals have found applications in astronomy, material science, medicine and biology \cite{blasquez2003efficient,guderlei2007algorithms,berchtold2007modelling,vogel2010quantification}, as well as voxel-based surface recognition \cite{yarotsky2017geometric}.

The additive property of Minkowski functional $\mathtt{Mv}$ for two convex sets A and B can be expressed as $\mathtt{Mv (A \cup B) = Mv (A) + Mv (B) - Mv (A \cap B)}$.This property, as well as discrete structure of the 3D voxel image, considerably simplifies the computation of Minkowski functionals for a rock sample dataset, since, for such objects, the procedure is reduced to enumeration of open voxels, faces, edges and vertices \cite{blasquez2003efficient}.

For efficient computation of the functionals, we utilized the method, described in \cite{blasquez2003efficient}, which uses binary decision diagrams. This method takes advantage of the local configuration around each added voxel.

\paragraph{Minkowski functionals for rescaled samples}

Another way to evaluate rock permeability using its voxel image is to include not only Minkowski functionals for the sample itself, but also functionals for a rescaled sample as an input to the machine learning algorithm \cite{srisutthiyakorn2016deep}. The intuition behind this technique is that, while Minkowski functionals retain some fine information about geometrical structure of the sample, calculating them for a rescaled sample could provide insights on the geometrical structure on a larger scale, offering a better mathematical description of the sample and its viscous flow properties.

A rescaled sample of magnitude $M$ is a voxel cube, the dimensions of which are $M$ times smaller. A given voxel is set to 1 (pore phase) if the average of voxels in the corresponding range in the initial sample is no less than a specified threshold. In this work, a threshold of 0.5, and magnitudes $M$ of 2, 5, 10 and 25 were used.

\paragraph{VGG-PCA descriptors}

VGG is the name for a family of deep convolutional neural networks (DCNNs) for 2D image recognition. They were introduced on ImageNet Challenge 2014 \cite{simonyan2014very}, and they marked the advent of the Deep Learning era. Not only could these networks provide accurate image classification and generalize well, but a pretrained network could be fine-tuned to a given problem, quickly achieving satisfactory performance in terms of some metric without additional increase of the dataset size and training time.

A common approach to fine-tuning is to remove several bottom layers from a pretrained DNN or CNN, exposing one of the dense layers, which are typically denoted as FC-1000 or FC-4096 (see the VGG architecture in \cite{simonyan2014very}). A number of layers is then added to the network, as appropriate to the specifics of the problem, and they are trained on the new data.

In our approach, we simply extract features from the FC-4096 layer for each 2D slice of the scan, represented as an image. After additional processing with PCA to reduce their dimensionality, these features are then used as inputs for the regression.
    
One important property of VGG network dense layers is that they retain a significant amount of image structure, and their output alone is frequently enough to correctly classify a given image or to process it in some other way. Although they are not interpretable, this set of values can provide much insight into composition, pattern distribution and other aspects of the image.
    
For the purposes of rock permeability prediction, we recovered descriptors of $100 \times 100 \times 1$ 2D layers, or, essentially, voxel rectangles, which compose a given $100 \times 100 \times 100$ sample in the dataset. To process the binary image, we converted all rock voxels to $(0, 0, 0)$ vectors in RGB code, and pore voxels to $(255, 255, 255)$. Accordingly, the voxel layers had to be resized to $224 \times 224$. The output of the second fully connected layer of size 4096 was used to recover the descriptors.
    
The feature vector of length 409600, obtained by concatenation of 100 layer descriptors, can then be interpreted as a descriptor of the sample as a whole, retaining information about the structures of individual layers and their position in the voxel cube through the placement of individual layers.
    
The enormous size of the vector makes it unsuitable for direct use as an input for the regression model. Instead, we used PCA to reduce dimensionality of the input vectors, making it possible to use conventional models without significant modifications. A principal component size of 1350 was used.
    
\subsection{Regression methods}

We used two regression methods to evaluate the predictive power of generated features: gradient regression trees (XgBoost) and deep neural networks (DNNs)
    
\paragraph{XgBoost}

XgBoost is a gradient boosting library \cite{xgboost}. It provides a powerful prediction model, consisting of numerous weak prediction models \cite{chen2016xgboost}. It is much used in Machine Learning due to its computation speed and interpretability of the results. We first found model hyperparameters, which yielded better results in terms of the $ABS_q$ metric (this metric is described in detail in section 5), by a grid search, i.e., by training the model with different hyperparameters and evaluating its performance on the hold-out validation subset.

The parameters used for all feature groups are described in Table 1.
    
\paragraph{Deep neural networks}

We used several deep multilayer perceptron (MLP) architectures to assess predictive power of generated features with neural networks. Final architecture and results are described in section 5.
    
\begin{table}[H]
  \centering
  \caption{XgBoost hyperparameters used} \label{tab:tab2}
  \begin{tabular}{ | l | c | r |}
    \hline
    \verb|learning_rate=0.05| & \verb|n_estimators=400| &  \verb|max_depth=5| \\ \hline
    \verb|min_child_weight=6| & \verb|gamma=0.1| & \verb|subsample=1| \\ \hline
    \verb|colsample_bytree=1| & \verb|reg_alpha=0.5| & \verb|reg_lambda=1| \\ \hline
  \end{tabular}
\end{table}

\section{End-to-End Regression}

We assessed the use of an end-to-end convolutional neural networks (CNN) modeling technique, which is commonly applied in image processing. We used a 2D CNN to carry out regression on individual 2D slices of a sample and a 3D CNN to process the samples as a whole.

\paragraph{2D convolutional neural networks}

CNNs are a class of deep feedforward artificial neural networks, which are most commonly applied to analyze visual imagery.

\begin{wrapfigure}{r}{0.5\textwidth}
    \centering
    \includegraphics[width=0.5\textwidth]{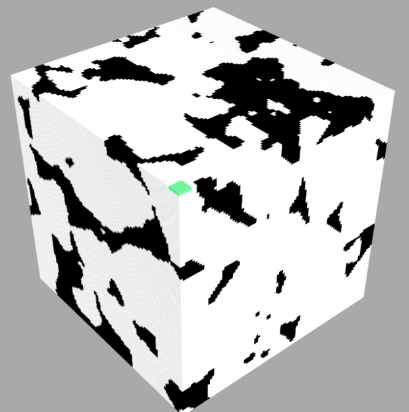}
    \caption{Receptive field for $5 \times 5$ filter of 2D CNN}
\end{wrapfigure}

Just like multilayer perceptrons, CNNs were inspired by biological structures, specifically, by the organization of the visual cortex of animals. Specific cortical neurons respond to corresponding stimuli only in a restricted region of the visual field, which is called the “receptive field” of those neurons \cite{matsugu2003subject}.

In artificial neural networks, this approach is realized by convolutional and pooling layers. A convolutional layer consists of a number of filters, each of which is trained to respond to a specific pattern. Filters are iterated over the input tensor, and the Hadamard product of filter weights and corresponding input values is calculated for each visited position. The sum of the elements in the resulting matrix is then passed on. The subset of input values, analyzed by a filter at a given step, is called the receptive field of the filter.

This generalization of a biological approach offers a number of advantages, such as shift-invariance and parameter sharing \cite{lecun2015deep}. But the most important point in the context of the task at hand is the emphasis given to the spatial structure of the data. As porosity depends greatly on the number and shapes of pores, the ability to treat input voxels which are close to each other differently from voxels which are far away is invaluable for estimating both local and global spatial structure of the rock sample.

CNNs are often used to analyze images, most commonly represented as 3D tensors, where the first two dimensions correspond to directions in the image, and the third dimension corresponds to color channels (red, green, blue) for a given pixel. In our work, we used 2D convolutions with 3D rock samples the same way as for images. But instead of vectors of color channels, we used vectors of voxel values on other layers: the three color channels are replaced by 100 values, each corresponding to a layer of the sample. After a series of other convolutions, max-pooling operations and fully connected layers, the predicted value for sample permeability is calculated.

\begin{wrapfigure}{l}{0.5\textwidth}
    \centering
    \includegraphics[width=0.5\textwidth]{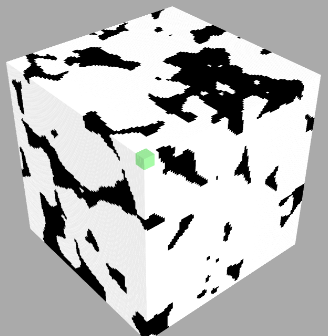}
    \caption{Receptive field for $5 \times 5 \times 5$ filter of 3D CNN}
\end{wrapfigure}

However, this approach has the significant disadvantage that each filter works with $5 \times 5 \times 1$ tensors, disregarding information from neighboring layers. This nuance is mitigated for 3D CNNs, where each filter examines the local region across all three dimensions of a 3D rock sample.

\paragraph{3D convolutional neural networks}

3D CNNs are based on the same idea as 2D CNNs, but 3D filters are used for convolutional layers. For purposes of permeability prediction, this allows a given filter to receive the local information for a given voxel not just from the same layer, but also from neighboring layers. As rock pores are three-dimensional, such an approach provides more practical information to each network unit.

\section{Model Evaluation and Results}

We compared the predictive power of feature sets and we also compared different prediction methods. The selection of methods used the criteria of interpretability and relatively straightforward modus operandi.

 	All prediction methods were compared with each other. For better interpretability of results, we used a special metric, denoted as $ABS_q:$
    
$$ABS_q = \frac{\frac{1}{n}\Sigma_{i=1}^n |y_i - \hat{y}_i|}{P_{99}(y)-P_{1}(y)}. $$

Here, $y_i$ denotes the true permeability value for sample $i$, $\hat{y}_i$ is a predicted permeability value for sample $i$, and $P_j(y)$ is the $j$th percentile of a true permeability histogram for a given cube. This metric is more informative than mean squared error. Conventional error does permit comparison of algorithmic approaches with each other, but provides zero information about how large the error is compared with variability of the data. Our approach takes account of such difference.

\paragraph{XgBoost}

The results for selected feature types and feature type combinations using the XgBoost approach are presented below. The last row of each table corresponds to the feature group combination, which yields the best result for the given method. Only error values below the 90th percentile were used to produce the charts, as each method generates strong outliers in terms of permeability. 

Here and below, VGG-PCA corresponds to introduced VGG-PCA descriptors, NET corresponds to rock sample network features, and MX corresponds to Minkowski functionals for an X-rescaled cube.

\begin{figure}[H]
    \centering
    \includegraphics[width=\textwidth]{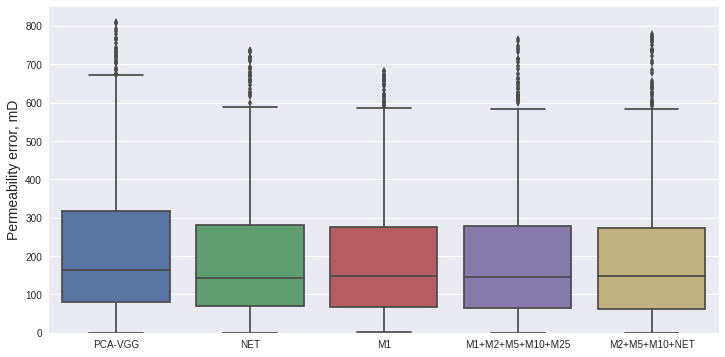}
    \caption{XgBoost permeability prediction errors ($<$90th percentile)}
\end{figure}

\begin{table}[H]
  \centering
  \caption{Evaluation of permeability prediction for selected feature group combinations} \label{tab:tab3}
  \begin{tabular}{ | l | c | }
    \hline
    Feature Groups Used & Validation $ABS_q$ \\ \hline
    VGG-PCA & 0.0451 \\ \hline
    NET & 0.0417 \\ \hline
    M1 & 0.0421 \\ \hline
    M1 + M2 + M5 + M10 + M25 & 0.0406 \\ \hline
    M2 + M5 + M10 + NET & 0.0368 \\ \hline
  \end{tabular}
\end{table}

To further improve the results, training and prediction were carried out with logarithms of permeability, and ABSq was computed for the exponent of prediction. This empirical technique has proved to give better results in some cases.

\begin{figure}[H]
    \centering
    \includegraphics[width=\textwidth]{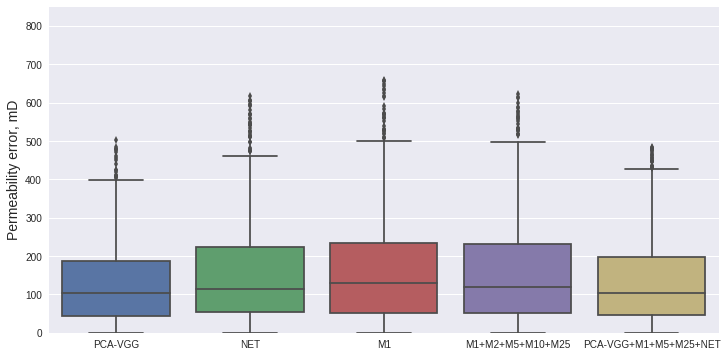}
    \caption{XgBoost logarithmic permeability prediction errors ($<$90th percentile)}
\end{figure}

\begin{table}[H]
  \centering
  \caption{Evaluation of logarithmic permeability for selected feature group combinations} \label{tab:tab4}
  \begin{tabular}{ | l | c | }
    \hline
    Feature Groups Used & Validation $ABS_q$ \\ \hline
    VGG-PCA & 0.0367 \\ \hline
    NET & 0.0372 \\ \hline
    M1 & 0.0391 \\ \hline
    M1 + M2 + M5 + M10 + M25 & 0.0370 \\ \hline
    VGG-PCA + M1 + M5 + M25 + NET & 0.0338 \\ \hline
  \end{tabular}
\end{table} 

The VGG-PCA features perform much better when used with logarithmic permeability. Despite not being able to provide the best results individually, we found that the VGG-PCA features are among the 25 top-scoring feature type combinations. It is interesting that the result using VGG-PCA features is much worse for usual permeability. However, examining the cause for that would require an excursion into VGG-16 architecture specifics, which goes beyond the scope of the present article.
These results should be regarded with a degree of caution, as, strictly speaking, they only correspond to the predictive power of these feature groups for a given sandstone sample, and only for the XgBoost model.

\paragraph{Deep neural networks}

\setlength\intextsep{0pt}
\begin{wraptable}[8]{r}{0.55\textwidth}
  \centering
  \caption{Used MLP architecture}\label{tab:tab5}
  \begin{tabular}{ | l |}
    \hline
    Dense (units=2048, activation="relu") \\ \hline
    Dense (units=2048, activation="relu") \\ \hline
    Dense (units=1024, activation="relu") \\ \hline
    Dense (units=512, activation="relu") \\ \hline
    Dense (units=256, activation="relu") \\ \hline
    Dense (units=1, activation=None) \\ \hline
  \end{tabular}
\end{wraptable} 

To evaluate the predictive power of feature groups paired with neural networks, we used three feature group combinations: VGG-PCA, M1 + M2 + M5 + M10 + M25 and VGG-PCA + M1 + M2 + M5 + M10 + M25. The number of considered feature group combinations was restricted in order to reduce time spent on training the models, since neural networks are much more computationally demanding.

We evaluated several straightforward multilayer perceptron (MLP) architectures for each feature group combination, and the best one was selected for comparison. It is presented in Table 4.

\begin{table}[H]
  \centering
  \caption{Evaluation of logarithmic permeability for selected feature group combinations} \label{tab:tab6}
  \begin{tabular}{ | l | c | }
    \hline
    Feature Groups Used & Validation $ABS_q$ \\ \hline
    VGG-PCA & 0.0287 \\ \hline
    M1 + M2 + M5 + M10 + M25 & 0.0441 \\ \hline
    VGG-PCA + M1 + M2 + M5 + M10 + M25 & 0.0384 \\ \hline
  \end{tabular}
\end{table}

The only difference between the best architectures is that Minkowski functionals seem to provide better results when a batch normalization layer is added before the output unit. In the following table we present $ABS_q$ value for all considered feature group combinations.

The addition of VGG-PCA descriptors to Minkowski functionals reduces prediction error. However, the best result is achieved when they are used separately from the other features. This is because the network was not given enough training time to nullify excess information coming from the Minkowski functionals, which introduced additional error.

The number of training epochs was limited to 50 for all tested architectures, batch size was set to 8, and the Adam optimizer with a learning rate of 0.001 was used. All remaining hyperparameters were set to default. Early stopping was used in order to determine the best validation score.

\paragraph{Convolutional neural networks}

Below we present the best performing 2D and 3D CNN architectures. Similarly to other approaches, logarithmic permeability was used as an output value. 

Used 2D CNN architecture is inspired by the VGG-16 network \cite{lecun2015deep}, which was used to compute VGG-PCA descriptors. We consider each sample to have 100 channels, each corresponding to an individual layer. The model was trained using the Adam optimizer with default parameters, for 20 epochs and a batch size of 32.

\begin{table}[H]
  \centering
  \caption{Best performing 2D CNN, $ABS_q$ = 0.0406} \label{tab:tab7}
  \begin{tabular}{ | c |}
    \hline
    2D Convolutional (filters=64, kernel\_size=3, activation="relu", padding="same") \\ \hline
    2D Convolutional (filters=64, kernel\_size=3, activation=“relu”, padding=“same”) \\ \hline
    2D Convolutional (filters=64, kernel\_size=3, activation=“relu”, padding=“same”) \\ \hline
    Max Pooling (pool\_size=(2, 2), strides=(2, 2)) \\ \hline
    2D Convolutional (filters=128, kernel\_size=3, activation=“relu”, padding=“same”) \\ \hline
    2D Convolutional (filters=128, kernel\_size=3, activation=“relu”, padding=“same”) \\ \hline
    2D Convolutional (filters=128, kernel\_size=3, activation=“relu”, padding=“same”) \\ \hline
    Max Pooling (pool\_size=(2, 2), strides=(2, 2)) \\ \hline
    2D Convolutional (filters=256, kernel\_size=3, activation=“relu”, padding=“same”) \\ \hline
    2D Convolutional (filters=256, kernel\_size=3, activation=“relu”, padding=“same”) \\ \hline
    2D Convolutional (filters=256, kernel\_size=3, activation=“relu”, padding=“same”) \\ \hline
    Max Pooling (pool\_size=(2, 2), strides=(2, 2)) \\ \hline
    2D Convolutional (filters=512, kernel\_size=3, activation=“relu”, padding=“same”) \\ \hline
    2D Convolutional (filters=512, kernel\_size=3, activation=“relu”, padding=“same”) \\ \hline
    2D Convolutional (filters=512, kernel\_size=3, activation=“relu”, padding=“same”) \\ \hline
    Max Pooling (pool\_size=(2, 2), strides=(2, 2)) \\ \hline
    Dense (1024, activation=“relu”) \\ \hline
    Dropout(0.5) \\ \hline
    Dense (512, activation=“relu”) \\ \hline
    Dropout(0.5) \\ \hline
    Dense (1, activation=None) \\ \hline
  \end{tabular}
\end{table}

Used 3D CNN architecture was inspired by VoxNet \cite{maturana2015voxnet}, which was initially used for object recognition. Compared with 2D convolutions, the application of 3D CNN is straightforward. The model was also trained using the Adam optimizer with default parameters for 20 epochs and a batch size of 32. Valid padding was used for all convolutional layers.

3D CNNs of similar structure have proven to be an efficient way of addressing the task of 3D shape retrieval \cite{notchenko2017large}, since they can learn efficient descriptors for 3D objects, which are bound to be effective for regression.

\begin{table}[H]
  \centering
  \caption{Best performing 3D CNN, $ABS_q$ = 0.0284} \label{tab:tab8}
  \begin{tabular}{ | c |}
    \hline
    3D Convolutional (filters=32, kernel\_size=5, strides=2, activation="relu") \\ \hline
    3D Convolutional (filters=32, kernel\_size=5, strides=2, activation="relu") \\ \hline
    Max Pooling 3D (pool\_size=(2, 2), strides=(1, 1)) \\ \hline
    3D Convolutional (filters=32, kernel\_size=3, strides=1, activation="relu") \\ \hline
    3D Convolutional (filters=32, kernel\_size=3, strides=1, activation="relu") \\ \hline
    Max Pooling 3D (pool\_size=(2, 2), strides=(1, 1)) \\ \hline
    Dense (128, activation=“relu”) \\ \hline
    Dense (64, activation=“relu”) \\ \hline
    Dense (1, activation=None) \\ \hline
  \end{tabular}
\end{table}

Below we present the boxplot of errors for neural network approaches for permeability prediction. Once, again, strong outliers lying above the 90th percentile of errors were not used.

\begin{figure}[H]
    \centering
  \includegraphics[width=\textwidth]{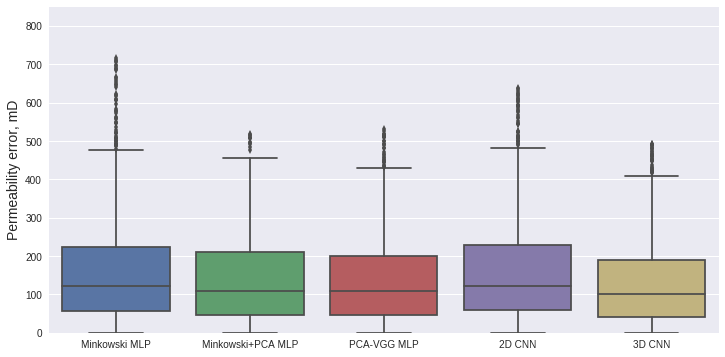}
    \caption{Neural network permeability prediction errors ($<$ 90th percentile)}
\end{figure}

\paragraph{Overall evaluation of the methods}

Below we present a comparison of the best results for each considered method. Overall, 3D convolutional neural networks proved to be superior both in terms of the $ABS_q$ metric and of error distribution. We would therefore recommend focusing on 3D convolutional neural networks for the development of data-driven permeability prediction models.

\begin{figure}[H]
    \centering
  \includegraphics[width=\textwidth]{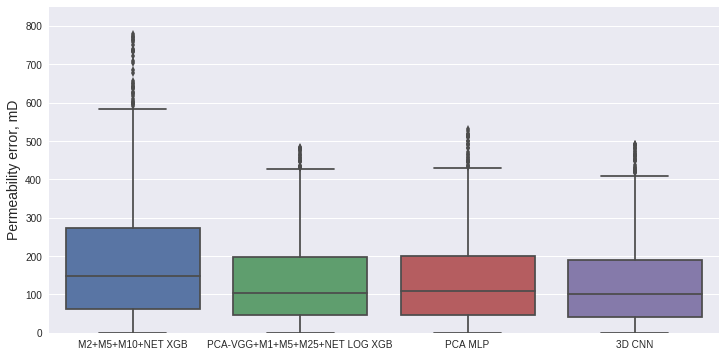}
    \caption{Permeability prediction error distribution for the best models ($<$ 90th percentile)}
\end{figure}

\begin{table}[H]
  \centering
  \caption{Evaluation of best models for each considered approach} \label{tab:tab7}
  \begin{tabular}{ | l | c | }
    \hline
    Model and Approach & $ABS_q$ \\ \hline
    M2+M5+M10+NET XGB & 0.0368 \\ \hline
    VGG-PCA+M1+M5+M25+NET LOG XGB & 0.0338 \\ \hline
    VGG-PCA MLP & 0.0287 \\ \hline
    3D CNN & 0.0284 \\ \hline
  \end{tabular}
\end{table}

\section{Conclusions and Discussion}

The results of this pilot study clearly demonstrate the significant potential of machine learning for image-based permeability prediction. It can be seen that the data-driven approach is a true game changer for Digital Rock technology because it is extremely fast and scalable. Moreover, the approach appears to be applicable not only to single-phase permeability prediction, but to prediction of more complex properties relevant to petrophysics, structural geology and field development. Such properties may include relative phase permeabilities, formation factor and resistivity, dielectric permittivity, elastic and geomechanical properties, NMR response and others. There are opportunities for enriching input data with information on mineral distribution, wettability and intergrain contacts to obtain the highest possible predictive power.
Clearly, there is no shortage of themes for future study and it should also be noted that the recent developments in Deep Learning are likely to enable prediction of the dynamics of fluid displacement, and not merely of the static characteristics of digitized rock samples of a single type. Assisted by a feature set containing information about pore fluids, this will represent a promising direction for applications of image-based Digital Rock in enhanced oil recovery work. We would emphasize that this direction will need to be developed together with physics-driven pore scale modeling \cite{koroteev2014direct}, since, without the physics-based models, there will be no actual data, on which to carry out training.

In future studies we plan to evaluate the use of more efficient DNN models and methods. These include modern approaches to constructing ensembles of regression models \cite{burnaev2013method} and special methods for the initialization of DNN parameters \cite{burnaev2016influence}.

We are also considering the adaptation of implemented models to other core types using multi-fidelity regression modeling methods. These are examined in \cite{zaytsev2017minimax,zaytsev2017large,burnaev2015surrogate}. Successful applications of such a technique include an application in aerodynamics, examined in \cite{belyaev2014building}

Another approach worth considering is to apply adaptive design of experiments, devised for industrial engineering problems, to both increase the efficiency of sensitivity analysis and improve utilization of the computational budget when generating a training sample \cite{burnaev2017efficient,burnaev2015adaptive}.

\section*{References}

\bibliography{mybibfile}

\end{document}